\documentstyle[prl,aps,twocolumn,epsf]{revtex}
\def \fra#1#2{{\textstyle {\frac{#1}{#2}}}}
\def \binom#1#2{{\textstyle {#1\choose #2}}}
\begin{document}

\twocolumn[\hsize\textwidth\columnwidth\hsize\csname
@twocolumnfalse\endcsname

\title{Maximum entanglement of quantum-dot systems with single excitation\\
in the spin van der Waals model}

\author{
A. Miranowicz,$^{(a,b,c)}$\cite{e-mail} \c{S}. K.
\"Ozdemir,$^{(a,b)}$ M. Koashi,$^{(a,b)}$ N. Imoto,$^{(a,b,d)}$
and Y. Hirayama$^{(a,d)}$}

\address{
$(a)$ CREST Research Team for Interacting Carrier Electronics,
Japan Science and Technology Corporation (JST)\\
$(b)$ School of Advanced Sciences, Graduate Univ. for Advanced
Studies (SOKEN), Hayama, Kanagawa 240-0193, Japan\\
$(c)$ Nonlinear Optics Division, Institute of Physics, Adam
Mickiewicz
University, 61-614 Pozna\'n, Poland\\
$(d)$ NTT Basic Research Laboratories, 3-1 Morinosato Wakamiya,
Atsugi, Kanagawa 243-0198, Japan}

\pagestyle{plain} \pagenumbering{arabic} \maketitle

\begin{abstract}
Time evolution of entanglement of $N$ quantum dots is analyzed
within the spin-$1/2$ van der Waals (or Lipkin-Meshkov-Glick) XY
model. It is shown that, for a single dot initially excited and
disentangled from the remaining unexcited dots, the maximum
bipartite entanglement can be obtained in the systems of
$N=2,\cdots,6$ dots only.

\vspace{5mm}
\end{abstract}
]

\vspace{4mm} {\bf Introduction.} Entanglement is one of the most
profound concepts of quantum mechanics. But only recently it has
been recognized as a possible resource for quantum information
processing including quantum computation, quantum teleportation,
quantum dense coding, or certain types of quantum key
distributions \cite{Preskill}.

To describe a bipartite entanglement of quantum-dot systems in a
pure state, given by $\hat{\rho}_{AB} =(|\psi\rangle\langle
\psi|)_{AB}$, we apply the von Neumann reduced entropy defined as
\cite{Ved97}
$$E[\hat{\rho}_{AB}]=
-{\rm Tr}\{\hat{\rho}_A\log_2 \hat{\rho}_A\} =-{\rm
Tr}\{\hat{\rho}_B\log_2 \hat{\rho}_B\}
$$
where $\hat{\rho}_A={\rm
Tr}_B\{\hat{\rho}_{AB}\}$ and $\hat{\rho}_B ={\rm Tr}
_A\{\hat{\rho}_{AB}\}$ are the reduced density matrices.

\vspace{4mm} {\bf Model.} \noindent We analyze entanglement in a
system of $N$ two-level quantum dots described by the Pauli spin
operators (${\hat{\sigma}}_n^x, {\hat{\sigma}}_n^y,
{\hat{\sigma}}_n^z$) for each ($n=1, \cdots , N$) lattice site by
assuming that each dot is coupled to all others with the same
strength independent of separation distance and nature. Such a
model is referred to as the spin van der Waals (SVW) \cite{Dek91},
Lipkin-Meshkov-Glick \cite{Lip65}, equivalent-neighbor or
infinitely-coordinated spin system. Some aspects of entanglement
in the SVW systems in different contexts have already been
analyzed in Refs. \cite{Mol99,Qui99,Koa00,Wan01}.

In particular, the quantum SVW system in the XY-like regime can be
defined by the interaction Hamiltonian~\cite{Dek91}:
$$\hat{H}_{{\rm int}} = \kappa
\sum_{n\neq m}\left({\hat{\sigma}}_n^{+} {\hat{\sigma}}_m^{-}+
{\hat{\sigma}}_n^{-} {\hat{\sigma}}_m^{+}\right).
$$
The creation and annihilation spin operators are given by $\sigma
_n^{\pm }={\hat{\sigma}}_n^x\pm {\hat{\sigma}}_n^y$, respectively.
Let's assume that the initial state describing a system of $M$
($M=0, \cdots , N$) dots excited and $N-M$ dots in the ground
state is given as $|\psi (0)\rangle = \{|1\rangle ^{\otimes M}\}_A
\{|0\rangle ^{\otimes (N-M)}\}_B$. Then, the solution of the
Schr\"{o}dinger equation of motion for the SVW model reads as
$$|\psi (t)\rangle
=\sum_{m=0}^{M^{\prime }}C_m^{NM}(t) {\{|1\rangle ^{M-m}|0\rangle
^m\}}_A\, {\{|1\rangle ^m|0\rangle ^{N-M-m}\}}_B
$$ where $M^{\prime }=\min (M, N-M)$. The states in curly brackets,
$\{|1\rangle ^{\otimes (M-m)}|0\rangle ^{\otimes m}\}$, denote a
sum of all possible [i.e., $\binom{M-m}m$] $M$-dot states with the
excitation number $(M-m)$.  The time-dependent superposition
coefficients, $C_m^{NM}(t)$, are given by
$$
C_m^{NM}(t) =\sum_{n=0}^{M^{\prime }}b_{mn}^{NM}\exp \Big\{ {\rm
i} [n(N+1-n)-M(N-M)]\kappa t\Big\}
$$
with
$$
b_{mn}^{NM} =\sum_{k=0}^m (-1)^k\binom mk\binom{N-2k} {M-k}^{-1}
\left\{\binom{N+1-2k}{n-k}-2\binom{N-2k}{n-k-1}\right\}
$$
where $\binom xy$ are binomial coefficients. In the following we
will focus on the evolution of the system of $N$ dots in which  a
single dot ($M=1$) is  initially excited.

\vspace*{0cm}
\begin{figure}
\hspace*{0cm} \epsfxsize=8cm \epsfbox{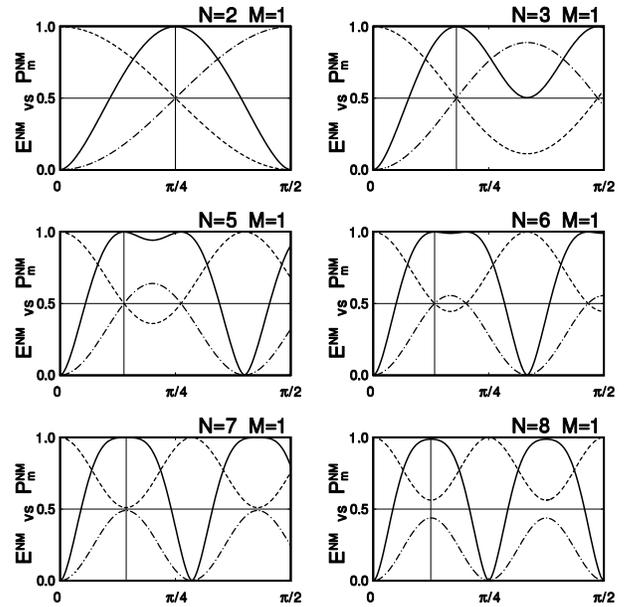} \vspace*{3mm}
\caption{ Evolution of the entanglement, $E^{NM}$ (solid curves),
and the Schmidt coefficients: $P^{NM}_1$ (dot-dashed curves) and
$P^{NM}_2$ (dashed curves) for systems of $N=2,\cdots,8$ quantum
dots with only one ($M=1$) of them initially excited. It is seen
that the maximum entanglement corresponds to the Schmidt
coefficients mutually equal or, in general, the least different.
} \label{fig2} \end{figure}

\newpage
\vspace*{-5mm}
\begin{figure}
\hspace*{0cm} \epsfxsize=8cm \epsfbox{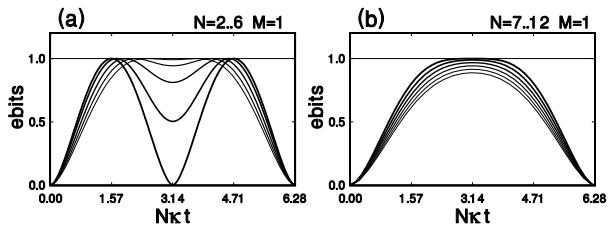} \vspace*{3mm}
\caption{ Entanglement as a function of the rescaled time $N\kappa
t$. It is seen that one ebit of entanglement can be achieved in
systems with the total number of dots ranging from two (thickest
curve in fig. (a)) up to six (thinnest curve in fig. (a)) only.
For $N>6$, the maximum entanglement decreases with increasing $N$.
The evolution of the systems with higher $N$ is marked by thinner
curves. } \label{fig1}
\end{figure}
\vspace*{-9mm}
\begin{figure}
\hspace*{5mm} \epsfxsize=6cm \epsfbox{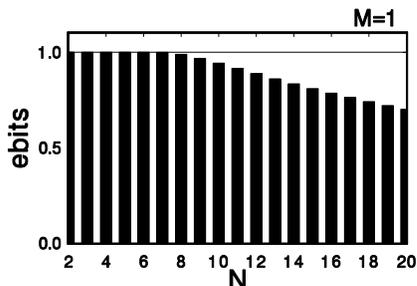} \vspace*{0cm}
\caption{ The maximum values of the entanglement $E^{N,1}$ as a
function of the total number $N$ of quantum dots in systems with
single excitation. On the scale of the figure, $\max_t
E^{7,1}(t)=0.9997$ does not differ from unity. However, for higher
values of $N$, a monotonous decrease of the maximum entanglement
is clearly visible.  } \label{fig5}
\end{figure}

{\bf Entanglement.}  The Schmidt coefficients for our solutions
are given by \cite{Mir01}
$$
P_m^{NM}(t)=\binom{M}m\binom{N-M}m |C_m^{NM}(t)|^2.
$$
They enable the direct calculation of the bipartite entanglement
via the Shannon entropy
$$E^{NM}(t)\equiv E[|\psi (t)\rangle\langle \psi (t)|]
=-\sum_m P^{NM}_m\log_2 P^{NM}_m.
$$
In the simplest nontrivial case, for $M=1$, the Schmidt
coefficients reduce to
$$P_1^{N,1}(t)=4\fra{(N-1)}{N^2}\sin^2(Nt/2)$$ and
$P_2^{N,1}(t)=1-P_1^{N,1}(t)$. The evolution of the Schmidt
coefficients, $P_m^{N,1}(t)$, and the entanglement, $E^{N,1}(t)$,
are depicted in Fig. 1.

The quantum-dot systems evolve into the maximum entangled state
for times given by
$$
0=\dot{E}^{NM}=2\fra{N-1}{N}\sin(N\kappa t)\\
\log_2\left[ \fra {N^2}{4(N-1)} \csc^2(\fra{N}{2}\kappa
t)-1\right]
$$
with the roots  $$\kappa t'=\fra 2N {\rm arccsc}(\fra 2N
\sqrt{2(N-1)})$$ and $\kappa t''=\frac{\pi}N$. We find that the
maximum of entanglement, equal to ${E}^{N,1}(t')=1$ ebit, can be
achieved for the evolution time  $t'$ for $N\le 6$ only. For
$N>6$, real solution for $t'$ does not exist. This conclusion can
also be drawn by analyzing Figs. 1 and 2.

Entanglement for $N>6$ reaches its maximum at the evolution times
$t''$. This maximum value, given by
\begin{eqnarray}
{E}^{N,1}(t'')&=&\fra 2{N^2} \big\{ N^2 \log_2 N-(N-2)^2\log_2
(N-2)
\nonumber \\
&&-2(N-1)\log_2 [4(N-1)]\big\}<1, \nonumber
\end{eqnarray}
monotonically decreases with increasing $N$ as clearly depicted in
Figs. 2 and 3.

\vspace{4mm} {\bf Conclusion.} We have studied evolution of
quantum dots in the spin van der Waals XY model. We have applied
its Schmidt decomposition to investigate information-theoretic
entropic properties and entanglement of the quantum dot ensembles.
Closer analysis of our solution shows that the quantum dots evolve
periodically into the bipartite entangled states. We have found
for the systems of $N$ dots with a single dot initially excited
and disentangled from the remaining unexcited $(N-1)$ dots that
the maximum bipartite entanglement can be observed in the systems
of $N=2,\cdots,6$ dots only.  It is worth noting that the same
``magic'' number $N=6$ has been found in a different context of
the equivalent-neighbor entangled webs in Ref. \cite{Koa00}.

\vspace{2mm} \noindent We thank J. Bajer, Y. X. Liu, H. Matsueda,
and T. Yamamoto for stimulating discussions.


\vspace{1mm} {\setlength{\fboxsep}{1pt}
\begin{center}
\framebox{\parbox{0.75\columnwidth}{%
\begin{center}
published in the {\em Technical Digest of CLEO/Pacific Rim 2001},
IEEE Pub. No. 01TH8557, 2001, vol. 2, p.  130-131
\end{center}}}
\end{center}

\end{document}